# A Watermarking Technique Using Discrete Curvelet Transform for Security of Multiple Biometric Features

**Rohit Thanki[a], Ved Vyas Dwivedi[b] & Komal Borisagar[c]**

[a]Ph.D.Research Scholar, E.C. Department, Faculty of Technology & Engineering, C. U. Shah University, Wadhwancity, India, Contact: rohitthanki9@gmail.com
[b]Pro Vice Chancellor, C. U. Shah University, Wadhwan City, India
[c]Assistant Professor, E.C. Department, Atmiya Institute of Technology & Science, Rajkot, India

The robustness and security of the biometric watermarking approach can be improved by using a multiple watermarking. This multiple watermarking proposed for improving security of biometric features and data. When the imposter tries to create the spoofed biometric feature, the invisible biometric watermark features can provide appropriate protection to multimedia data. In this paper, a biometric watermarking technique with multiple biometric watermarks are proposed in which biometric features of fingerprint, face, iris and signature is embedded in the image. Before embedding, fingerprint, iris, face and signature features are extracted using Shen-Castan edge detection and Principal Component Analysis. These all biometric watermark features are embedded into various mid band frequency curvelet coefficients of host image. All four fingerprint features, iris features, facial features and signature features are the biometric characteristics of the individual and they are used for cross verification and copyright protection if any manipulation occurs. The proposed technique is fragile enough; features cannot be extracted from the watermarked image when an imposter tries to remove watermark features illegally. It can use for multiple copyright authentication and verification.

## 1. INTRODUCTION

The rapidly use of internet and growth in multimedia data communication over the internet has created a problem of copyright authentication and protection. For prevention against copyright authentication and protection, Digital Rights Management (DRM) system provides one of the solutions. DRM refers to a range of access control mechanism used for copyright, digital data access [1]. Digital watermarking is one of the technologies used in DRM systems to provide copyright protection and authentication for digital data. Digital watermarking can hide important information into the digital data, such as digital images, audio and video [1 – 3].
Biometrics features are referred as behavioral and physiological characteristics of an individual. These biometrics features of an individual are used for an individual verification and authentication. These biometrics features are unique for every individual. These biometrics features used as watermark in digital watermarking technique to provide unique identity of an individual, to improve ownership and protection of multimedia data.

## 1.1. Importance of Biometric Features as Watermark Information

The watermarking technique embeds a name, logo or text information of copyrighted individual into the host digital data which can be text information, digital image, video and audio [3]. There are some limitations of these types of watermarks such as less important information, less related to a copyrighted individual for authentication. These types of watermarks are easy to available, easy to tamper and reproduced. The convention watermarking technique does not provide validate copyright authentication because of digital data are watermarked using a particular text information or logo by imposter [1].

Recently, researchers have been using biometric features of individuals in watermarking technique to improve the copyright authentication and protection availability of conventional watermarking. This new idea is divided into two primary types such as biometric watermarking and multibiometric watermarking.

## 1.2. Biometric and Multibiometric Watermarking

The biometric characteristics are a watermark, which is used for watermarked any digital data. By embedding biometric watermark in the host data, it converted into such information about individual identity which can be hardly modified by the imposter. Biometric characteristics such as fingerprint, handwritten signature, face and iris are widely used for individual authentication. These biometric characteristics are easy to available, accessible and widely used in world wide. The combination of biometric data with the watermarking technique, then it is called biometric watermarking [1, 2].

Where biometric data are taken as a host medium and watermarked with biometric watermark features, then it is called multibiometric watermarked. The watermarking technique is insertion biometric data into other biometric data called as multibiometric watermarking technique [4, 5, 6]. For example, minutiae points of fingerprint are embedded in to face image as watermark. The multibiometric watermarking technique can use for multiple verification of an individual or protection of biometric data over non-secure transmission medium. This concept arises for the protection of biometric data against spoofing attack at the system database of biometric because the imposter can't access multiple biometric features where one biometric feature is embedded into another biometric feature. The biometric watermarking is more secure compared to conventional watermarking because every individual have uniequ biometric features.

Jain and his research team are for use first time biometric features as watermark and give the concept of watermarking of biometric features [4, 5, 6, 7, 8]. Jain and Uludag proposed copyright protection of digital data based on biometric watermark features such as the minutiae points of fingerprint and facial features of individual [4, 5, 7, 8]. They have described a watermarking technique which used fingerprint features as watermark to secure facial features of an individual to improve overall security of biometric systems. Other related work available in the literature for fingerprint [7, 9, 10, 11, 12], iris [13, 14, 15, 16, 17, 18], voice [1, 17, 19, 20], face [7, 9, 13, 20, 21, 22] and signature [1, 23] as a watermark data inserted into other biometric features for multiple verification of an individual or protection of biometric data over non-secure transmission medium.

Digital watermarking has reached its extreme level, but still research required in biometric watermarking. The related work to biometric watermarking in which fingerprint [7, 23, 24], iris [26, 27], voice [28] and signature [29, 30] as a watermark data inserted into digital data such as images, video and 3D model is available in the literature.

### 1.3. Multiple Watermarking

This type of watermarking approach can be used for multiple data protection or single data protection may be used several times. For example, a first watermark can be used to embed name and other information of the individual, a second watermark for ID information and a third watermark for fingerprinting. The multiple watermarking techniques can be holding multiple copyright information. According to the various applications, watermarking techniques having different properties, e.g., for copyright protection, robustness property required or for copyright authentication, fragility property required. The multiple watermarking techniques can be divided into below types [1, 31].
1. Composite Watermarking: With this type of technique, all multiple watermarks are combined as a single watermark which is subsequently embedded in one single embedding step. Figure 1 shows the composite watermarking approach for multiple watermark protection.
2. Segmented Watermarking: In this type of technique, host data are divided into different segments and watermarks information are inserted into these segments.

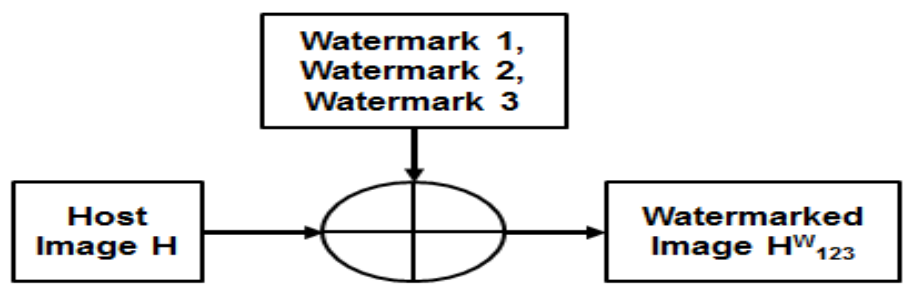

Figure 1. Composite Multiple Watermarking

3. Successive Watermarking: In this type of technique, watermarks are embedded one after the other. This approach is also known as re-watermarking. Figure 2 shows the successive watermarking approach for multiple watermark protection.

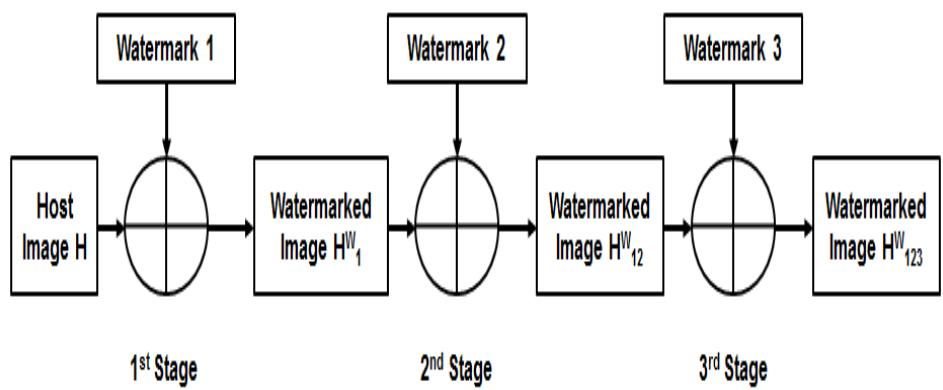

## 2. REFERENCE WATERMARKING TECHNIQUE

There is one reference watermarking technique available in the literature. Rege and her research team (Inamdar & Rege, 2014) give dual watermarking technique in which the authors have embedded more than one watermark biometric information in different wavelet coefficients of host image. In this technique, authors purposed a watermarking technique which embeds multiple biometric information. Authors are described that PN sequences do not have any information about individuals. Therefore, authors are using various watermark biometric features such as voice, face and signatures which is depending for owner characteristics.

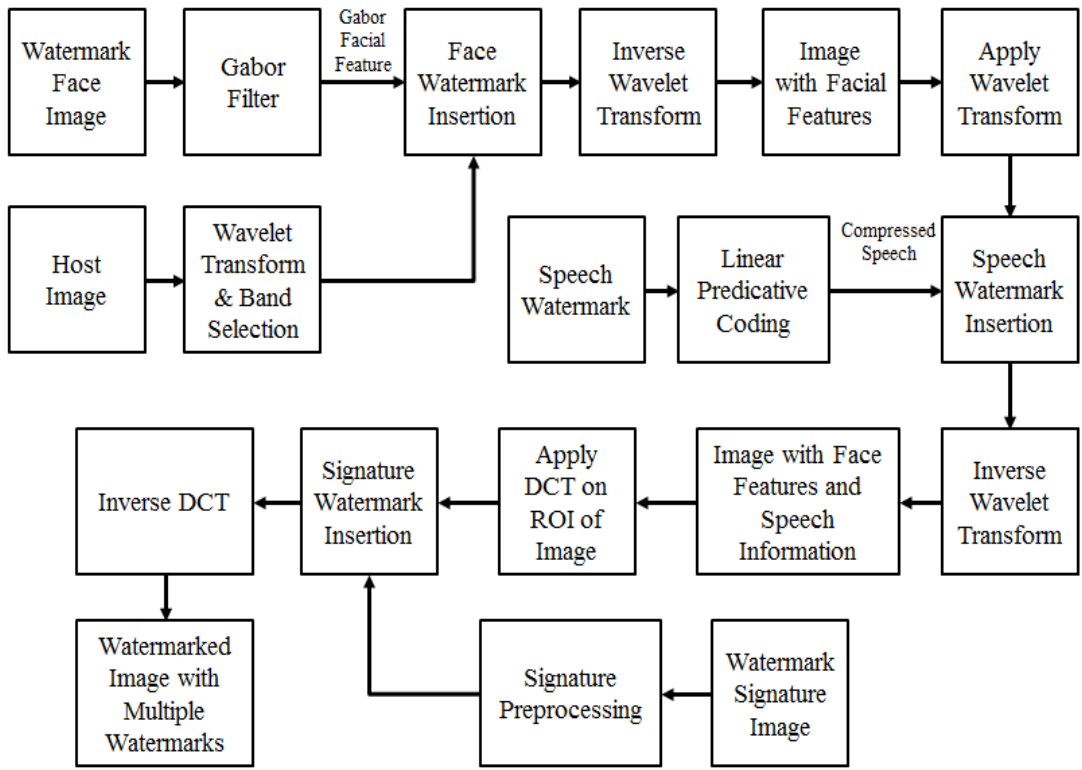

Figure 3. Block Diagram of Reference Watermarking Technique [1]

This technique based on Discrete Wavelet Transform (DWT) and Discrete Cosine Transform (DCT) decomposition. In this technique, information of two watermarks biometric is embedded invisibly in various wavelet subbands of host image and one watermark biometric information is embedded visibly in DCT coefficients of watermarked image where two biometric images are already embedded. The multiple watermark biometrics are embedded as depicted in Figure 3. This technique based on successive multiple watermarking. The main drawback of this technique is that, as the embedding procedure, various transform such as DCT and DWT is required. The watermark embedding procedure is given below:

- Take a face image as watermark. Then applied Gabor filter on it to extract facial features of the user.
- Take a host image and apply wavelet transform on it to convert into various subbands of wavelet coefficients. Then chose one subband of wavelet coefficients of host image where Gabor facial features are inserted to get a watermarked image with facial features.
- Take a speech of the user as watermark and apply Linear Predicative Coding (LPC) on it to get compressed speech features of the user.
- Then applied wavelets transform on watermarked image and get various subbands of wavelet coefficients of watermarked image.
- The compressed speech features are inserted into one subband of wavelet coefficients of watermarked image to get a watermarked image with facial features and speech information.
- Take a signature of the user as a watermark which is embedded visible in watermarked image which have facial features and speech information.
- For watermark signature embedding, apply Discrete Cosine Transform (DCT) on ROI of watermarked image which have facial features and speech information to get DCT coefficients of watermarked image.
- Then watermark signature is embedded into DCT coefficients of watermarked image. Apply inverse DCT on modified DCT coefficients to get the final version of watermarked image.
- The final version of watermarked image having invisible facial features and speech information with visible signature on it.

## 3. PROPOSED WATERMARKING TECHNIQUE

In this paper, a multiple watermarking technique is proposed for embedding multiple biometric features into single host data. The biometric watermarks, individual's fingerprint, iris, face and offline signature are used. The mostly watermarking techniques use a noise sequence as a watermark and a binary decision, whether the digital data is watermarked or not. This noise sequence does not have significant value. Significant motivations for using features of fingerprint, iris, face and signature as watermark data because of there are unique for every individual, easily available and widely used for authentication. Also, existing watermarking techniques, different image transforms such as DCT and DWT is required for multiple watermark embedding. These are these limitations and reasons which motivated us to propose this watermarking technique.

The block diagram for proposed technique is shown in Figure 4. We proposed a composite watermarking technique for multiple watermark insertion. This proposed technique embeds fingerprint features in the curvelet coefficients of host image. Then embeds iris features in other selected curvelet coefficients of host image. Then facial features are embedded into another selected curvelet coefficients of host image. Finally, signature features embedded in other selected curvelet coefficients of host image. The selection of curvelet coefficients depends on size of host image and size of watermark biometric features. The watermark biometric features can be extracted using Shen-Castan edge detection (ISEF) algorithm [32] and Principal Component Analysis (PCA). The watermarking technique proposed such that it is recovered all watermarks information in a single stage.

This watermarking technique divided into four different procedures: (1) Feature extraction of biometric watermarks (2) Multiple biometric watermark insertion (3)

Multiple biometric watermark extraction (4) Comparison of recovered biometric watermark features with original watermark features for individual authentication.

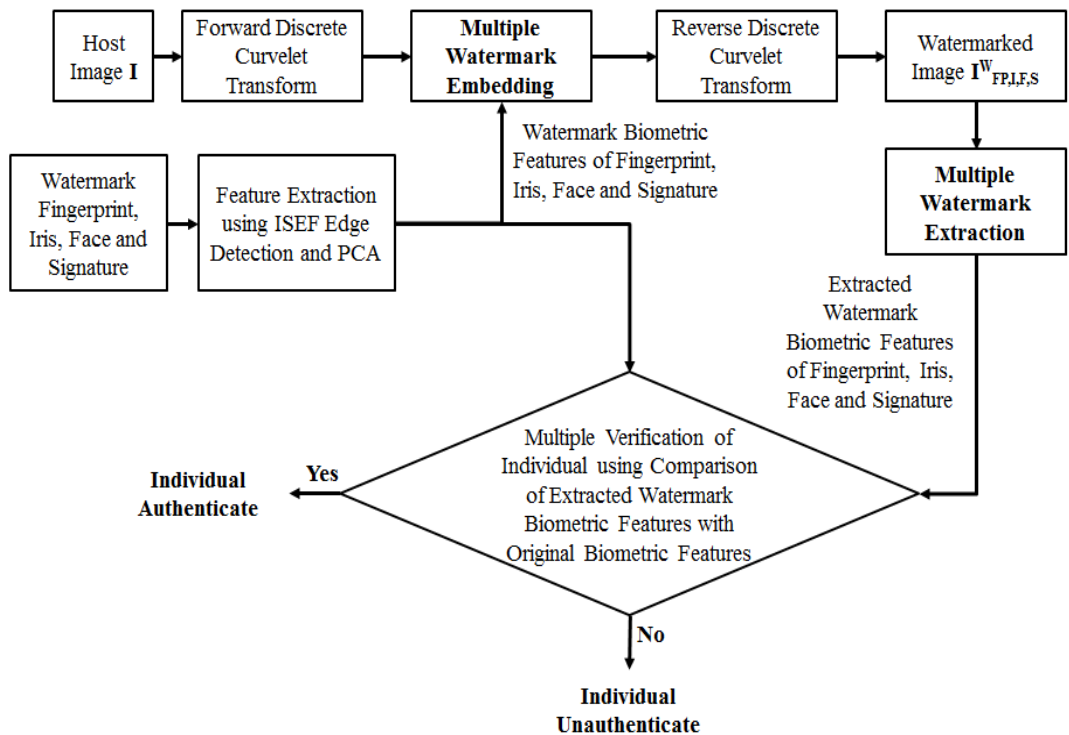

Figure 4. Block Diagram of Proposed Multiple Watermarking Technique

### 3.1. Feature Extraction of Biometric Watermarks

Various approaches such as core point detection, minutiae maps, orientation maps, Gabor feature maps used for extraction of fingerprint features; Gabor filter, zero crossing of 1D wavelets and Haar encoding using for extraction of iris features; Eigen face using Principal Component Analysis (PCA), Fisher faces using Linear Discrimination Analysis (LDA), Independent Component Analysis (ICA) used for extraction of facial features; Fixed point arithmetic, horizontal and vertical point splitting used for extraction of signature features. These all techniques used for generation biometric features which can be used for individual authentication or verification.

In this paper, we have generated biometric features of fingerprint, iris, face and signature characteristics using ISEF edge detection algorithm and Principal Component Analysis (PCA). These biometric features are used as watermark information and individual authentication. These biometric features are compared with extracted biometric features for individual authentication at detector side.

### 3.1.1. ISEF Edge Detection Algorithm

The Shen – Castan is introduced novel edge detection algorithm based on infinite symmetric exponential filter (ISEF) [32]. The authors agree with Canny about the general form of the edge detector a convolution with a smoothing kernel followed by a search for edge pixels. The steps of the ISEF edge detection algorithm are given below when this algorithm used for edge extraction of biometric image. Figure 5 shows the result of the ISEF edge detector on the biometric image.

Step 1: Recursive Filtering in X Direction
Step 2: Recursive Filtering in Y Direction
Step 3: Apply Binary Laplacian Technique
Step 4: Apply Non Maxima Suppression
Step 5: Find the Gradient
Step 6: Apply Hysteresis Thresholding
Step 7: Apply Thinning

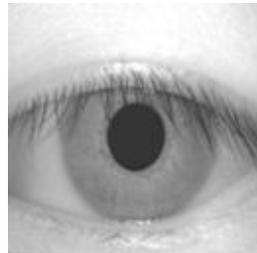 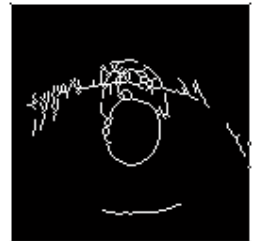

(a) Original Biometric Image   (b) Edge Detection of Biometric Image
Using ISEF Edge Detection

Figure 5. ISEF Edge Detector Performance on Biometric Image

### 3.1.2. Principal Component Analysis

Dimensionality reduction is one of the most common processes while biometric features are extracted from its characteristics. Principal Component Analysis (PCA) is one of the approaches used for dimensional reduction of biometric features. PCA is a statistical approach which transforms correlated variables into a set of values of linear uncorrelated variables called principal components. In this paper, after getting the edges of the watermark biometric image, PCA is applied to the edges to get the features of biometric image which is shown in Figure 6.

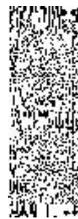

Figure 6. Biometric Features after Application of PCA on Edges of Biometric Image

### 3.2. Multiple Biometric Watermark Insertion

In this stage, multiple biometric features information taken as watermark information which is inserted into the host image. In this proposed algorithm, embeds biometric features in the host image by utilizing various curvelet coefficients introduced by the Fast Discrete Curvelet Transform (FDCT).

### 3.2.1. Curvelet Decomposition of Host Image

There are two types of curvelet transform such as continuous curvelet transform and discrete curvelet transform available in the literature. Discrete curvelet transform is used most of image processing applications such as compression, watermarking and edge detection [33]. The frequency wrapping based discrete curvelet transform technique is easy to implement, less computation time and easy to understand compared to the USFFT technique [33, 34]. Therefore, frequency wrapping based curvelet transform technique is used in many image processing applications. When the frequency wrapping based curvelet transform [33, 34]capplied to an image, then the image is converted into low frequency coefficients and high frequency coefficients.

The curvelet decomposition of image by frequency wrapping based curvelet transform is decomposed image into various cells which is denoted as C. For example, host image with a size of 512×512 pixels are decomposed using fast discrete curvelet transform into different cells, such as C {1, 1} to C {1, 6}. Where, C {1, 1} is low frequency curvelet coefficients while other C {1, 2}, C {1, 3}, C {1, 4}, C {1, 5}, C {1, 6} is high frequency curvelet coefficients. The curvelet decomposition of the host image is shown in Figure 7. In this paper, we have chosen various curvelet coefficients of cell C {1, 5}. The selection of curvelet coefficients depends on size of host image and size of watermark biometric features.

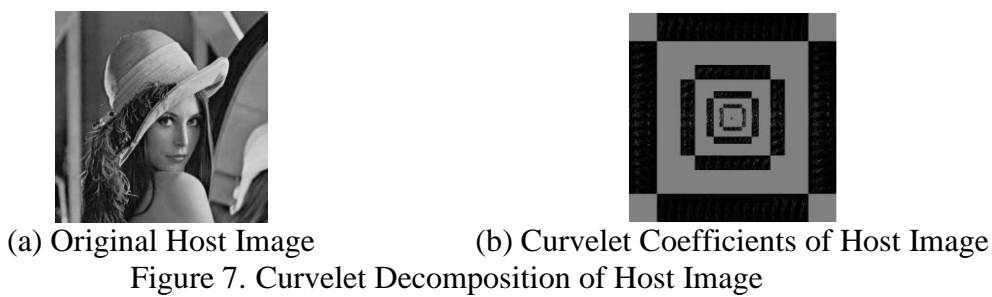

(a) Original Host Image  (b) Curvelet Coefficients of Host Image

Figure 7. Curvelet Decomposition of Host Image

### 3.2.2. Multiple Biometric Watermark Embedding

1. Take a fingerprint, iris, face and signature of the individual as a watermark.
2. Then extracted biometric features using ISEF edge detection and PCA. These biometric features are used as watermark information and which is denoted as $F_{FP}$, $F_I$, $F_F$ and $F_S$. Then compute the size of biometric features.
3. Take host image and compute the size of the image.
4. Then the forward frequency wrapping based fast discrete curvelet transform is applied to host image convert into various curvelet coefficients. The reason behind choosing in high frequency coefficients C {1, 5} is that the size of high frequency curvelet coefficients C {1, 5} is the same size of biometric watermark features.
5. Then various cells of high frequency curvelet coefficients C {1, 5} of the host image are modified according to values of biometric features and gain factor using Cox equation [35].

$$C_{Watermarked}\{1,5\}\{1,2\} = C_{Curvelet\_Coefficients}\{1,5\}\{1,2\} * (1 + k \times F_{FP}) \quad (1)$$

$$C_{Watermarked}\{1,5\}\{1,4\} = C_{Curvelet\_Coefficients}\{1,5\}\{1,4\} * (1 + k \times F_I) \quad (2)$$

$$C_{Watermarked}\{1,5\}\{1,5\} = C_{Curvelet\_Coefficients}\{1,5\}\{1,5\} * (1 + k \times F_F) \quad (3)$$

$$C_{Watermarked}\{1,5\}\{1,8\} = C_{Curvelet\_Coefficients}\{1,5\}\{1,8\} * (1 + k \times F_S) \quad (4)$$

Where $C_{Watermarked}$ {1, 5} = various modified high frequency curvelet coefficients; $C_{Curvelet\_Coefficients}$ {1, 5} = original high frequency curvelet coefficients; K = gain factor, $F_{FP}$ = features of fingerprint watermark, $F_I$ = features of iris watermark, $F_F$ = features of face watermark, $F_S$ = features of signature watermark.

6. Applied reverse frequency wrapping based fast discrete curvelet transform on modified curvelet coefficients with another unmodified curvelet coefficient to get watermarked biometric image.

### 3.3. Multiple Biometric Watermark Extraction

1. Take a watermarked image and apply the frequency wrapping based fast discrete curvelet transform (FDCT) on it to convert into various curvelet coefficients such as $C_W\{1, 1\}, C_W\{1, 2\}, C_W\{1, 3\}, C_W\{1, 4\}, C_W\{1, 5\}, C_W\{1, 6\}$.
2. Take an original host image and apply the frequency wrapping based fast discrete curvelet transform (FDCT) on it to convert into various curvelet coefficients such as $C\{1, 1\}, C\{1, 2\}, C\{1, 3\}, C\{1, 4\}, C\{1, 5\}, C\{1, 6\}$.
3. Extracted sparse measurements of a watermark biometric image using the reverse procedure of embedding.

$$RF_{FP} = \frac{\left(\frac{C_{Watermarked}\{1.5\}\{1,2\}}{C_{Curvelet\_Coefficients}\{1,5\}\{1,2\}} - 1\right)}{k} \quad (5)$$

$$RF_{I} = \frac{\left(\frac{C_{Watermarked}\{1.5\}\{1,4\}}{C_{Curvelet\_Coefficients}\{1,5\}\{1,4\}} - 1\right)}{k} \quad (6)$$

$$RF_{F} = \frac{\left(\frac{C_{Watermarked}\{1.5\}\{1,5\}}{C_{Curvelet\_Coefficients}\{1,5\}\{1,5\}} - 1\right)}{k} \quad (7)$$

$$RF_{S} = \frac{\left(\frac{C_{Watermarked}\{1.5\}\{1,8\}}{C_{Curvelet\_Coefficients}\{1,5\}\{1,8\}} - 1\right)}{k} \quad (8)$$

Where $C_{Watermarked}\{1, 5\}$ = various watermarked high frequency curvelet coefficients; $C_{Curvelet\_Coefficients}\{1, 5\}$ = original high frequency curvelet coefficients; K = gain factor, $RF_{FP}$ = recovered features of fingerprint watermark, $RF_I$ = recovered features of iris watermark, $RF_F$ = recovered features of face watermark, $RF_S$ = recovered features of signature watermark.

### 3.4. Comparison of Recovered Biometric Watermark Features with Original Watermark Features for Individual Authentication

1. After recovering biometric watermark features at detector side and then a comparison between recovering biometric watermark features and original biometric watermark features performed for decision about individual authentication.
2. For a decision about individual authentication, quality measures such as Structural Similarity Index Measure (SSIM) is used [36]. SSIM is find similarity between two features.
3. Here we have performed multiple authentication of the individual using below procedure where we have applied an average sum of SSIM values of fingerprint, iris, face and signature features for individual authentication.

$$S_1 = SSIM(F_{FP}, RF_{FP})$$
$$S_2 = SSIM(F_I, RF_I)$$
$$S_3 = SSIM(F_F, RF_F) \quad (9)$$
$$S_4 = SSIM(F_S, RF_S)$$
$$S = \frac{S_1 + S_2 + S_3 + S_4}{4}$$

Where $S_1$ = similarity between fingerprint watermark features, $S_2$ = similarity between iris watermark features, $S_3$ = similarity between face watermark features, $S_4$ = similarity between signature watermark features, S = average similarity between multiple watermark features

4. We have created two hypotheses for individual authentication is given below:
   - **Hypothesis 1:** Individual is authenticate if $S > \tau$
   - **Hypothesis 2:** Individual is unauthenticate if $S < \tau$
     Where S = average similarity value between multiple watermark features, τ = decision threshold.

## 4. EXPERIMENTAL RESULTS

In this proposed technique, multiple biometric watermarks embed into a single host image, and then testing is carried out to checking the effect of various watermarks on host image, quality of watermarked image and performance of individual authentication. The perceptual quantitative measure used for quality check of watermarked image is PSNR. Structural Similarity Index Measure (SSIM) is used for quality check for biometric features and individual authentication. Figure 8 shows the host images used for testing and fingerprint, iris, face and signature image used as a watermark. The size of host image is 512×512 pixels and a size of watermark image is 128×128.

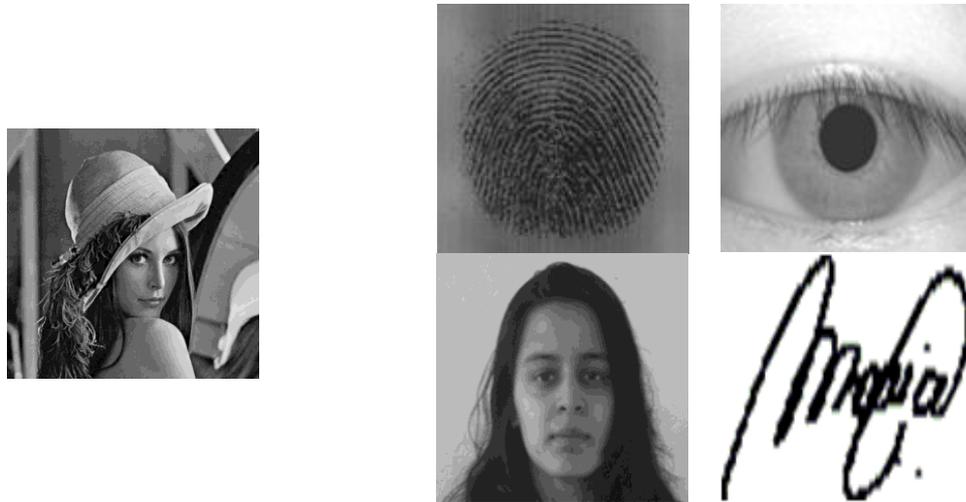

Figure 8. Standard Lean Image as Host image and Multiple Biometric Characteristics as Watermark

For generation of biometric watermark feature, we first apply ISEF edge detection algorithm on biometric watermark characteristics to get an edge of the biometric watermark image. Then applied PCA on the edges of the biometric watermark image and get the features of biometric watermark. These biometric features are used as watermark information. The size of fingerprint watermark features is 43×128, size of iris watermark features is 44×128, size of face watermark features is 44×128, size of signature watermark features is 43×128 which is equal to size of curvelet coefficients C {1, 5} {1, 2}, C {1, 5} {1, 4}, C {1, 5} {1, 5} and C {1, 5} {1, 8} of the host image respectively. Figure 9 shows the features of fingerprint watermark, iris watermark, face watermark and signature watermark.

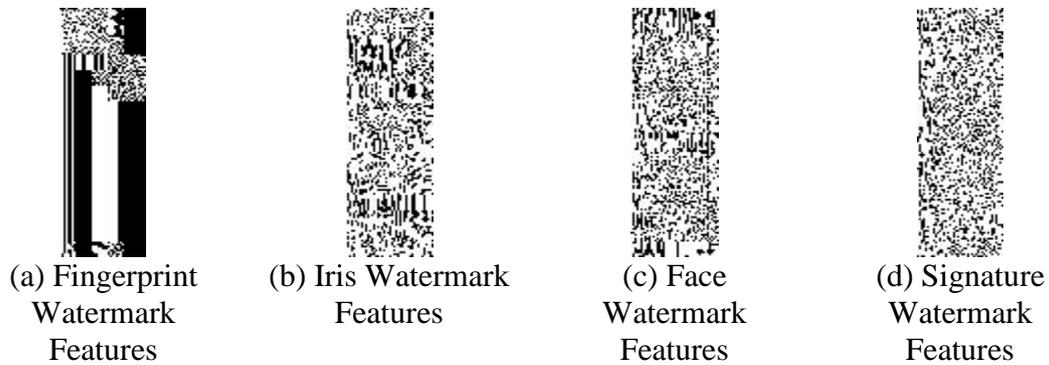

(a) Fingerprint Watermark Features    (b) Iris Watermark Features    (c) Face Watermark Features    (d) Signature Watermark Features

Figure 9. Biometric Watermark Features after Application of ISEF Edge Detection and PCA

These biometric watermark features are embedding into various curvelet coefficients of host image using equation 1 to 4 to get watermarked image (Figure 10 (a)). For watermark insertion, gain value of 0.02 is chosen. At extraction, reverse procedure of watermark feature insertion is followed. Figure 10 (b to e) shows recovered biometric watermark features at detector side.

For testing of proposed watermarking technique, various watermarking attacks such as JPEG compression, addition of different noise, apply different filter and geometric attacks like cropping. Table 1 summarized the PSNR value between original image and watermarked image and SSIM values between multiple biometric features and recovered multiple biometric features under consideration of the above attacks.

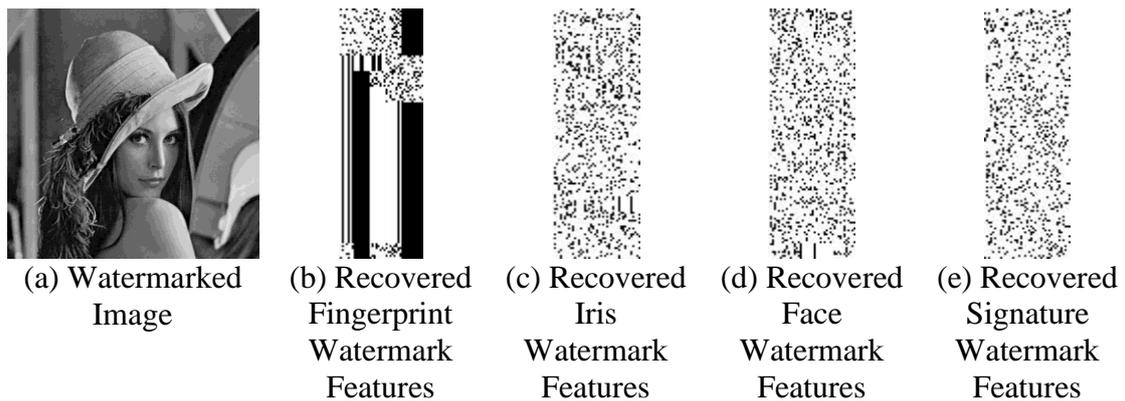

(a) Watermarked Image    (b) Recovered Fingerprint Watermark Features    (c) Recovered Iris Watermark Features    (d) Recovered Face Watermark Features    (e) Recovered Signature Watermark Features

Figure 10. Watermarked Image and Recovered Biometric Watermark Features

Table 1. Values of Quality Measures for Proposed Watermarking Technique

| Attacks | PSNR (dB) | $S_1$ | $S_2$ | $S_3$ | $S_4$ |
|---|---|---|---|---|---|
| No Attack | 42.44 | 1.00 | 1.00 | 1.00 | 1.00 |
| JPEG Compression (Q = 80) | 35.53 | 0.27 | 0.35 | 0.39 | 0.51 |
| JPEG Compression (Q = 70) | 34.68 | 0.04 | 0.22 | 0.38 | 0.37 |
| Gaussian Noise ( μ =0, σ=0.001) | 29.72 | 0.13 | 0.19 | 0.22 | 0.19 |
| Salt & Pepper Noise (Noise Density = 0.005) | 28.03 | 0.07 | 0.12 | 0.15 | 0.16 |
| Speckle Noise (Variance = 0.004) | 30.77 | 0.13 | 0.20 | 0.20 | 0.20 |
| Median Filter (size = 3 × 3) | 36.43 | 0.13 | 0.14 | 0.22 | 0.19 |
| Mean Filter (size = 3 × 3) | 31.08 | 0.09 | 0.02 | 0.00 | 0.02 |
| Gaussian Low Pass Filter (size = 3 × 3) | 38.28 | 0.33 | 0.52 | 0.54 | 0.47 |
| Histogram Equalization | 16.79 | 0.37 | 0.52 | 0.62 | 0.57 |
| Cropping | 30.03 | 0.52 | 0.52 | 0.60 | 0.10 |

SSIM values between biometric watermark features are used for decision about individual authentication. The threshold for individual decision is set 0.90. This proposed watermarking technique is fragile against various watermarking attacks based on SSIM value shown in Table 1. The individual is unauthenticate when various watermarking attacks on watermarked image because the average SSIM value of multiple biometric watermark features is less than 0.90 which is indicated in Table 2. This situation is indicated that this proposed watermarking technique can be provided copyright authentication of digital data based on multiple biometric watermark features.

In the watermark insertion process, the gain factor is multiplied with biometric watermark features, and then embedded in the curvelet coefficients of host image. Therefore, the gain factor has an effect on the watermarked image. Table 3 shows the effect of the gain factor changes on the PSNR value of the watermarked image. The gain factor value range is 0.005 to 0.05 for this proposed watermarking technique because when increase gain factor above 0.05 then this technique does not fulfilled the requirement of HVS property of watermarking.

Table 2. Average SSIM Value for Decision about Individual Authentication

| Attacks | Average SSIM Value (S) | Decision about Individual Authentication |
|---|---|---|
| No Attack | 1.00 | Authenticate |
| JPEG Compression (Q = 80) | 0.38 | Unauthenticate |
| JPEG Compression (Q = 70) | 0.25 | Unauthenticate |
| Gaussian Noise ( μ =0, σ=0.001) | 0.18 | Unauthenticate |
| Salt & Pepper Noise (Noise Density = 0.005) | 0.12 | Unauthenticate |
| Speckle Noise (Variance = 0.004) | 0.18 | Unauthenticate |
| Median Filter (size = 3 × 3) | 0.17 | Unauthenticate |
| Mean Filter (size = 3 × 3) | 0.03 | Unauthenticate |
| Gaussian Low Pass Filter (size = 3 × 3) | 0.47 | Unauthenticate |
| Histogram Equalization | 0.52 | Unauthenticate |
| Cropping | 0.44 | Unauthenticate |

Table 3. Effect of Gain Factor on Watermark Insertion

| Gain Factor | PSNR (dB) |
|---|---|
| 0.005 | 48.47 |
| 0.010 | 42.44 |
| 0.015 | 38.92 |
| 0.020 | 36.42 |
| 0.025 | 34.49 |
| 0.030 | 32.90 |
| 0.035 | 31.56 |
| 0.040 | 30.40 |
| 0.045 | 29.38 |
| 0.050 | 28.46 |
| 0.055 | 27.64 |

The proposed watermarking technique is compared with existing watermarking techniques available in literature with various features and parameters are summarized in Table 4. The existing watermarking techniques available in the literature are robust against various watermarking attacks while this proposed watermarking technique is fragile against attacks. The PSNR value shows that proposed watermarking technique is outperformed compared to existing watermarking techniques available in the literature.

Table 4. Comparison of Proposed Watermarking Technique with Existing Watermarking Techniques in the Literature

| Features | Mark Technique (2007) [31] | Inamdar Technique (2014) [1] | Proposed Technique |
|---|---|---|---|
| Type of Multiple Watermarking | Successive | Successive | Composite |
| Used Watermark Information | PN Sequences | Biometric Trait | Biometric Features |
| Used Image Transform for Watermark Embedding | Discrete Cosine Transform (DCT) | Discrete Wavelet Transform (DWT) | Fast Discrete Curvelet Transform (FDCT) |
| PSNR value | 40 dB | 35.18 dB | 42.44 dB |

## CONCLUSION

In this paper, we have proposed a new approach for authentication of watermarked image. We have proposed multiple biometric feature embedding technique by using ISEF edge detection, PCA and Fast Discrete Curvelet Transform (FDCT) for individual recognition and copyright authentication. We have presented various approaches for multiple biometric watermarking techniques for copyright protection and authentication. We have proposed that in multiple watermarking, using various transform coefficients of curvelet, where watermark embedding can be reduced complexity. Invisible watermark biometric features provide authentication of copyright ownership of digital data. The experimental results show that our approach is performed better than existing watermarking approaches available in the literature.